\documentclass[aps,prl, superscriptaddress, showpacs,twocolumn,nofootinbib,nobalancelastpage]{revtex4-1}
\usepackage{amsmath}
\usepackage{graphicx}
\usepackage{bm}
\usepackage{hyperref}
\usepackage{xcolor}
\usepackage{multirow}
\usepackage{amssymb}
\usepackage{soul}

\newcommand{\eEDM}{{\em e}EDM}

\begin{document}
\title{$\mathcal{T,P}$-odd effects in the LuOH$^+$ cation.}

\begin{abstract}
   The LuOH$^+$ cation is a promising system to search for manifestations of time reversal and spatial parity violation effects. Such effects in LuOH$^+$ induced by the electron electric dipole moment $e$EDM and the scalar-pseudoscalar interaction of the nucleus with electrons, characterized by $k_s$ constant, in LuOH$^+$ are studied. The enhancement factors, polarization in the external electric field, hyperfine interaction, rovibrational structure are calculated. The study is required for the experiment preparation and extraction of the $e$EDM and $k_s$ values from experimental data.
\end{abstract}

\author{Daniel E. Maison}
\email{daniel.majson@gmail.com, maison\_de@pnpi.nrcki.ru}
\affiliation{Petersburg Nuclear Physics Institute named by B.P.\ Konstantinov of National Research Center ``Kurchatov Institute'' (NRC ``Kurchatov Institute'' - PNPI), 1 Orlova roscha mcr., Gatchina, 188300 Leningrad region, Russia}
\homepage{http://www.qchem.pnpi.spb.ru    }

\author{Leonid V.\ Skripnikov}
\email{skripnikov\_lv@pnpi.nrcki.ru,leonidos239@gmail.com}
\affiliation{Petersburg Nuclear Physics Institute named by B.P.\ Konstantinov of National Research Center ``Kurchatov Institute'' (NRC ``Kurchatov Institute'' - PNPI), 1 Orlova roscha mcr., Gatchina, 188300 Leningrad region, Russia}
\affiliation{Saint Petersburg State University, 7/9 Universitetskaya nab., St. Petersburg, 199034 Russia}

\author{Gleb Penyazkov}
\email{glebpenyazkov@gmail.com}
\affiliation{Petersburg Nuclear Physics Institute named by B.P.\ Konstantinov of National Research Center ``Kurchatov Institute'' (NRC ``Kurchatov Institute'' - PNPI), 1 Orlova roscha mcr., Gatchina, 188300 Leningrad region, Russia}
\affiliation{Saint Petersburg State University, 7/9 Universitetskaya nab., St. Petersburg, 199034 Russia}

\author{Matt Grau}
\email{mgrau@odu.edu}
\affiliation{Department of Physics, Old Dominion University, Norfolk, VA 23529}

\author{Alexander N. Petrov}
\email{petrov\_an@pnpi.nrcki.ru}
\affiliation{Petersburg Nuclear Physics Institute named by B.P.\ Konstantinov of National Research Center ``Kurchatov Institute'' (NRC ``Kurchatov Institute'' - PNPI), 1 Orlova roscha mcr., Gatchina, 188300 Leningrad region, Russia}
\affiliation{Saint Petersburg State University, 7/9 Universitetskaya nab., St. Petersburg, 199034 Russia}

\maketitle

\section{Introduction}

For a long time, it was supposed, that the laws of physics should satisfy the conditions of the invariance with respect to charge conjugation ($\mathcal{C}$), spatial parity ($\mathcal{P}$) and time reversion ($\mathcal{T}$) symmetries. However, in the second half of XXth century it was experimentally confirmed, that both $\mathcal{P}$- and combined $\mathcal{CP}$-symmetries are violated in weak interactions. 
According to the $\mathcal{CPT}$ theorem, violation of $\mathcal{CP}$ is equivalent to violation of $\mathcal{T}$ symmetry.
The search for the new manifestations of violation of these symmetries
is one of the topical sections of modern theoretical and experimental physics \cite{Safronova:18}. 
For example, $\mathcal{CP}$-violation is of great interest for cosmology and astrophysics, since it is one of three necessary conditions of baryogenesis \cite{sakharov1967violation}.

One of the approaches to search for simultaneous violation of $\mathcal{T}$ and $\mathcal{P}$ symmetries ($\mathcal{T,P}$-violation) is the determination of elementary particles electric dipole moments (EDMs) \cite{whitepaper}. For example, the recent refinement of the neutron EDM upper constraint \cite{abel2020_neutronEDM} led to the updated constraint of the quantum chromodynamics (QCD) parameter $\bar{\theta}$ \cite{swallows2013techniques}. In addition, the electron electric dipole moment ($e$EDM) can be also used as indicator of $\mathcal{T,P}$-violation in the standard model (SM) and the physics beyond it. The latest result of the ACME collaboration in 2018 allowed one to obtain the strongest constraint on $e$EDM $|d_e| \lesssim 1.1\cdot 10^{-29}\ e\cdot\textrm{cm}$ \cite{ACME:18}. It overcame the previous constraints~\cite{ACME:14a,Cornell:2017} by almost an order of magnitude, and led to strong restrictions for various SM extensions. According to estimates within the SM, $e$EDM value is ten orders of magnitude lower \cite{Khriplovich:97, Yamaguchi:2021}, so there is still room for more precise experiments to search for new physics before encountering the SM background.

Recently it was suggested to perform $e$EDM search experiments using linear triatomic molecules \cite{kozyryev2017precision, isaev2017laser}. 
In Ref. \cite{kozyryev2017precision}, it was noted, that, due to the $l$-doubling effect in the first excited bending mode $\nu_2=1$, the linear triatomic molecules, such as YbOH, can be completely polarized by relatively weak electric field, $\sim 100 \textrm{V/cm}$ \footnote{However, later it was pointed out, that for some states the maximal polarization is only 50\% and no one state get 100\% polarization \cite{Petrov:2022}.}.  More importantly, these molecules can be successfully cooled~\cite{augenbraun2020laser}. These facts allow one to increase experimental sensitivity to $e$EDM and other $\mathcal{T,P}$-odd effects. 

As it was demonstrated in Ref.~\cite{Cornell:2017}, molecular cations can also be used for the $e$EDM determination. While experiments with ions suffer from reduced count rates, they are advantaged by long interrogation times afforded by the ion trap. The constraint $|d_e| \lesssim 1.3 \cdot 10^{-28} e\cdot\textrm{cm}$ obtained in this experiment is only an order of magnitude weaker, than the current one. This fact demonstrates, that the updated $e$EDM restrictions can possibly be obtained in ion trap experiments.

Combining these two ideas, it was suggested to consider the  LuOH$^+$ molecular ion for $\mathcal{T,P}$-odd effects search ~\cite{Maison:20a}, as it can be formed from  Lu atomic ions (which can be laser-cooled ~\cite{kaewuam:2018aa}), and once formed it can be sympathetically during an experiment cooled by co-trapped atomic ions. It has an electronic structure similar to YbOH. However, as it has been shown~\cite{Maison:20a}, the LuOH$^+$ molecular cation can be  even more sensitive to the nuclear $\mathcal{CP}$-violation effects, than YbOH, owing to the large electric quadrupole moment of $^{176}$Lu.

In a polar molecule with an atom of a heavy element, the $\mathcal{T,P}$-violating energy shift associated with $e$EDM and the scalar-pseudoscalar nucleus-electron interaction characterized by the dimensionless coupling constant $k_s$ reads
\begin{equation}
\Delta E_{\mathcal{P},\mathcal{T}}= P\left( E_{\rm eff}  d_e + E_{\rm s} k_s\right).
\label{shift}
\end{equation}
The $\mathcal{T,P}$-violating energy shift induced by $e$EDM in a molecule is determined by the following Hamiltonian~\cite{MartenssonPendrill:1987,Lindroth:89}:
\begin{eqnarray}
  H_d^{{\rm eff}}= d_e\sum_a  2i  c\gamma^0_a\gamma_a^5\bm{p}_a^2,
 \label{Wd2}
\end{eqnarray}
index $a$ runs over electrons (as in all equations below), $\bm{p}$ is the momentum operator for an electron and $\gamma^0$ and $\gamma^5=-i\gamma_0\gamma_1\gamma_2\gamma_3$ are the Dirac matrices, defined according to Ref.~\cite{Khriplovich:91} 
\footnote{Note that in the literature, there are two common definitions of $\gamma_5$ which differ by the sign.}. 
For a linear molecule this interaction can be characterized by the molecular constant $W_d$:
\begin{equation}
\label{matrelem}
W_d = \frac{1}{\Omega}
\langle \Psi|\frac{H_d}{d_e}|\Psi
\rangle.
\end{equation}
%
In these designations effective electric field introduced in Eq.~(\ref{shift}) acting on the electron electric dipole moment is $E_{\rm eff}=W_d|\Omega|$.
Another considered source of $\mathcal{T}, \mathcal{P}$-violation is the scalar-pseudoscalar nucleus-electron interaction given by the following Hamiltonian (see \cite{Ginges:04}, Eq.~(130), and also \cite{SF78}):
\begin{eqnarray}
  H_{\rm s}=i\frac{G_F}{\sqrt{2}}Zk_{\rm s}\sum_a \gamma^0_{a}\gamma^5_{a}\rho_N(\textbf{r}_a),
 \label{Htp}
\end{eqnarray}
where $G_F$ is the Fermi-coupling constant, $Z$ is the heavy nucleus charge, $\rho_N(\textbf{r})$ is the nuclear density normalized to unity and  $\mathbf{r}$ is the electron radius-vector with respect to the heavy atom nucleus under consideration.
This interaction is characterized by the molecular parameter $W_{T,P}$:
\begin{equation}
\label{WTP}
W_{T,P} = \frac{1}{\Omega}
\langle \Psi|\frac{H_{\rm s}}{k_{\rm s}}|\Psi
\rangle, 
\end{equation}
or $E_{\rm s}=W_{T,P} |\Omega|$.

 It is well known that for diatomics (like ThO, HfF$^{+}$) the polarization $P$ in Eq.~(\ref{shift}) 
 \footnote{ $P$ in Eq. (\ref{shift}), in general, is not equal to the mean value of the projection of unit vector $\hat{z}$ along molecular axis on direction of the external electric field.}
 smoothly approaches unity for small laboratory electric fields due to the existence of $\Omega$-doublet structure \cite{Cossel:12}. In Ref.~\cite{Petrov:2022} we showed, however, that $l$-doubling structure is in general different from $\Omega$-doubling, and that polarization tends to approach $|P| =0.5$ value for molecules (like YbOH,  LuOH$^+$, RaOH, etc.) with Hund's case $b$ coupling scheme. The final value depends on the value of the $l$-doubling, spin-rotation constant and hyperfine interaction.
 
Knowing the enhancement coefficients  $E_{\rm eff}$, $E_{\rm s}$ and $P$ one may extract the value of the constants $d_e$ and $k_s$ from the measured energy shift.
 
 To populate the required $\nu_2=1$ level in experiments one needs to know the bending vibrational energy levels structure. Therefore, these calculations are also performed in the present paper. Up to now there is no corresponding experimental information.

\section{Electronic structure calculation details}

The electronic structure calculations were performed within the relativistic coupled cluster approach. It is based on the exponential ansatz of wave function
\begin{equation}
    \Psi = \exp\left(\hat{T}\right) \Phi,
\end{equation}
where $\Phi$ is the electronic wave function in the Dirac-Hartree-Fock approximation, $\Psi$ is the wave function with electronic correlation taken into account, and $\hat{T}$ is the excitation cluster operator. It can be written as the following series:
\begin{equation}
    \hat{T} = \hat{T}_1 + \hat{T}_2 + \hat{T}_3 + \dots,
\end{equation}
with operators $\hat{T}_k$ defined as:
\begin{equation}
    \hat{T}_k = \frac{1}{k!} \sum\limits_{\substack{i_1<i_2<\dots<i_k \in \textrm{occ} \\b_1<b_2<\dots<b_k \in \textrm{virt}}}
    t_{i_1 i_2\dots i_k}^{b_1 b_2 \dots b_k} 
    \hat{a}_{b_1}^{\dagger} \hat{a}_{b_2}^{\dagger} \dots \hat{a}_{b_k}^{\dagger} 
    \hat{a}_{i_1} \hat{a}_{i_2} \dots \hat{a}_{i_k}.
\end{equation}
Indexes $i_{\dots}$ and $b_{\dots}$ label occupied and virtual electronic states, respectively. Coefficients $t_{\dots}^{\dots}$ are scalar variables to be defined, also called cluster amplitudes. In the present study, we exploit the relativistic coupled cluster approach with single and double cluster amplitudes (CCSD) and coupled cluster approach with single, double and perturbative triplet amplitudes (CCSD(T)) \cite{Crawford:00, Bartlett1991}. The former one exploits the approximation
$$
    \hat{T} \approx \hat{T}_1 + \hat{T}_2,
$$
and the latter one includes also calculation of the energy correction due to $\hat{T}_3$~\cite{Bartlett:2007}.

For calculation of the potential energy surface we have used the AE3Z(Lu) $\oplus$ aug-cc-PVTZ-DK(O,H) basis sets \cite{gomes:2010, Dunning:89, Kendall:92}. At the coupled cluster stage we have excluded $1s..3d$ electrons of Lu from the correlation treatment and set virtual orbitals energy cutoff equal to $70\ \textrm{Hartree}$.
The calculations of magnetic dipole hyperfine structure constants on hydrogen nucleus (see below), ion-frame electric dipole moment with respect to mass center and spin-rotational constant have been performed within the same procedure, with replacement the basis set aug-cc-PVTZ-DK on hydrogen by the AAE4Z basis set \cite{Dyall:2016}. 
Calculations of parameters of $\mathcal{T,P}$-violation effects in Eq.~(\ref{shift}) and electric quadrupole hyperfine constant induced by Lu nucleus have been performed at the CCSD(T) level of theory and using the AE3Z(Lu) $\oplus$ aug-cc-PVTZ-DK(O,H) basis as in Ref.~\cite{Maison:20a}. In these calculations all electrons were included in correlation calculation and virtual energy cutoff was set to 11000 $\textrm{Hartree}$. The contribution of high-energy orbitals in correlation calculation has been extensively analyzed in Refs.~\cite{Skripnikov:17a, Skripnikov:15a}.
Calculations of properties have been performed for the equilibrium geometry parameters determined in the present paper.

Molecular relativistic CCSD(T) calculations were carried out using the {\sc dirac} \cite{DIRAC19,Saue:2020} program package. Calculation of the property integrals were performed within the code developed in Refs.~\cite{Skripnikov:16b,Skripnikov:17b}.

\section{Rovibrational levels calculation details}
\begin{figure}[h]
\centering
\includegraphics[width=0.5\textwidth]{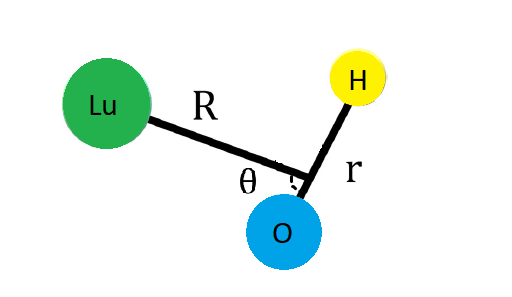}
\caption{Jacobi coordinates for LuOH$^+$ molecule.}
\label{Fig1}
\end{figure}

Following Ref. \cite{Petrov:2022}  we present our Hamiltonian in molecular reference frame as
\begin{equation}
{\rm \bf\hat{H}} = {\rm \bf\hat{H}}_{\rm mol} + {\rm \bf\hat{H}}_{\rm hfs} + {\rm \bf\hat{H}}_{\rm ext},
\label{Hamtot}
\end{equation} 
where
\begin{equation}
\hat{\rm H}_{\rm mol}=\frac{(\hat{\bf J} -\hat{\bf J}^{e-v} )^2}{2\mu R^2}+\frac{(\hat{\bf J}^{v})^2}{2\mu_{OH}r^2}+ V(\theta)
\label{Hmolf}
\end{equation}
is the molecular Hamiltonian as it is described in Ref.~\cite{Petrov:2022},
$\mu$ is the reduced mass of the Lu-OH system, $\mu_{OH}$ is the reduced mass of the OH, $\hat{\bf J}$ is the total electronic, vibrational, and rotational
angular momentum, $\hat{\bf J}^{e-v} = \hat{\bf J}^{e} + \hat{\bf J}^{v}$ is the electronic-vibrational momentum, $\hat{\bf J}^{e}$, is the electronic momentum, $\hat{\bf J}^{v}$ is the vibrational momentum,
$R$ is the distance between Lu and the center mass of OH, $r$ is OH bond length
and $\theta$ is the angle between OH  and the axis ($z$ axis of the molecular frame) directed from Lu to the OH center of mass. The condition $\theta=0$ corresponds to the linear configuration where the O atom is between Lu and H ones. $R$, $r$ and $\theta$ are the so called Jacobi coordinates, see Fig.~(\ref{Fig1}).

 In the current work, we have fixed $R$ and $r$,  to their equilibrium values obtained in the electronic structure calculations.
In this approximation we neglect the influence of the stretching $\nu_1$ (associated with R) and OH ligand $\nu_3$ (associated with r)  modes but nevertheless
take into account the bending ones (associated with $\theta$) with fixed $R,r$.
$V(\theta)$ is the potential energy curve obtained in the electronic structure calculations.

\begin{equation}
\begin{aligned}
 {\rm \bf\hat{H}}_{\rm hfs} = 
- { g}_{\rm H} {\bf \rm I^H} \cdot
 \sum_a\left(\frac{\bm{\alpha}_{2a}\times \bm{r}_{2a}}{r_{2a}^3 }\right) + \\
-{ g}_{\rm Lu}{\mu_{N}} {\bf \rm I^{Lu}} \cdot
\sum_a\left(\frac{\bm{\alpha}_a\times \bm{r}_{1a}}{{r_{1a}}^3}\right) \\
-e^2 \sum_q (-1)^q \hat{Q}^2_q({\bf \rm I^{\rm Lu}}) \sum_a \sqrt{\frac{2\pi}{5}}\frac {Y_{2q}(\theta_{1a}, \phi_{1a})}{{r_{1a}}^3}
\end{aligned}
\end{equation}
is the hyperfine interaction of electrons with Lu and H nuclei,
${ g}_{\rm Lu}$ and ${ g}_{\rm H}$ are the
 g-factors of the lutetium and hydrogen nuclei, $\bm{\alpha}_a$
 are the Dirac matrices for the $a$-th electron, $\bm{r}_{1a}$ and $\bm{r}_{2a}$ are their
 radius-vectors in the coordinate system centered on the Lu and H nuclei,
 $\hat{Q}^2_q({\bf \rm I^{\rm Lu}})$ is the quadrupole moment operator for $^{175}$Lu nucleus, $I^{\rm Lu}=7/2$, $I^{\rm H}=1/2$ are nuclear spins for $^{175}$Lu and H.

The Stark Hamiltonian
\begin{equation}
 {\rm \bf\hat{H}}_{\rm ext} =   -{ {\bf D}} \cdot {\bf E}
\end{equation}
describes the interaction of the molecule with the external electric field, and
{\bf D} is the dipole moment operator.

Wavefunctions, rovibrational energies and hyperfine structure were obtained by numerical diagonalization of the Hamiltonian (\ref{Hamtot})
over the basis set of the electronic-rotational-vibrational-nuclear spins wavefunctions
\begin{equation}
 \Psi_{\Omega m\omega}P_{lm}(\theta)\Theta^{J}_{M_J,\omega}(\alpha,\beta)U^{\rm H}_{M^{\rm H}_I}U^{\rm Lu}_{M^{\rm Lu}_I}.
\label{basis}
\end{equation}
Here 
 $\Theta^{J}_{M_J,\omega}(\alpha,\beta)=\sqrt{(2J+1)/{4\pi}}D^{J}_{M_J,\omega}(\alpha,\beta,\gamma=0)$ is the rotational wavefunction, $\alpha,\beta$ correspond to azimuthal and polar angles of the $z$ axis,
 $U^{\rm H}_{M^{\rm H}_I}$ and $U^{\rm Lu}_{M^{\rm Lu}_I}$ are the hydrogen and lutetium nuclear spin wavefunctions, $M_J$ is the projection of the molecular (electronic-rotational-vibrational) angular momentum $\hat{\bf J}$ on the lab axis, 
 $\omega$ is the projection of the same momentum on $z$ axis of the molecular frame,
 $M^{\rm H}_I$ and $M^{\rm Lu}_I$ are the projections of the nuclear angular 
momenta of hydrogen and lutetium on the lab axis, $P_{lm}(\theta)$ is the associated Legendre polynomial,
$l$ is the vibration angular momentum and $m$ is its projection on the molecular axis, 
$\Psi_{\Omega m\omega}$ is the electronic wavefunction (see Ref. \cite{Petrov:2022} for details).

In this  calculation functions with $\omega - m = \Omega = \pm 1/2$, $l=0-30$  and $m=0,\pm 1, \pm 2$, $J=1/2,3/2,5/2$  were included to the basis set (\ref{basis}).
The ground vibrational state $\nu_2=0$ corresponds to $m=0$,
the first excited bending mode $\nu_2=1$ to $m=\pm 1$, the second excited bending mode has states with $m=0, \pm2$ etc. A common designation $\nu_2^m$ for vibrational levels will be used below.

Provided that the {\it electronic-vibrational} matrix elements are known, the matrix elements of ${\rm \bf\hat{H}}$ between states in the basis set (\ref{basis}) can be calculated with help of the angular momentum algebra \cite{LL77, Petrov:2022} mostly in the same way as for the diatomic molecules \cite{Petrov:11}.

The required matrix elements associated with $^{175}$Lu nucleus magnetic hyperfine interaction 
\begin{multline}
A_{ \parallel} = -\frac{g_{\rm Lu}}{\Omega} \times\\
   \langle
   \Psi_{\Omega m\omega}P_{lm} |\sum_a\left(\frac{\bm{\alpha}_{1a}\times
\bm{r}_{1a}}{r_{1a}^3}\right)
_z|\Psi_{\Omega m \omega}P_{l'm}\rangle \\
= 8142~\delta_{ll'} {~ \rm MHz},
\end{multline}

\begin{multline}
A_{ \perp} = -{g_{\rm Lu}} \times\\
   \langle
   \Psi_{\Omega=1/2m\omega}P_{lm} |\sum_a\left(\frac{\bm{\alpha}_a\times
\bm{r}_{1a}}{r_{1a}^3}\right)
_+|\Psi_{\Omega=-1/2 m \omega-1}P_{l'm}\rangle \\
= 7864~ \delta_{ll'} {~ \rm MHz},
\end{multline}
 were taken from Ref. \cite{Maison:20a}.

Matrix elements associated with the hyperfine interaction induced by the H nucleus magnetic 
\begin{multline}
A_{ \parallel} = -\frac{g_{\rm H}}{\Omega} \times\\
   \langle
   \Psi_{\Omega m\omega}P_{lm} |\sum_a\left(\frac{\bm{\alpha}_a\times
\bm{r}_a}{r_a^3}\right)
_z|\Psi_{\Omega m \omega}P_{l'm}\rangle \\
= 
2.4
\delta_{ll'} {~ \rm MHz},
\end{multline}

\begin{multline}
A_{ \perp} = -{g_{\rm H}} \times\\
   \langle
   \Psi_{\Omega=1/2m\omega}P_{lm} |\sum_i\left(\frac{\bm{\alpha}_i\times
\bm{r}_i}{r_i^3}\right)
_+|\Psi_{\Omega=-1/2 m \omega-1}P_{l'm}\rangle \\
= 
-0.9
\delta_{ll'} 
{~ \rm MHz},
\end{multline}
dipole moment operator 
\begin{equation}
   \langle
   \Psi_{\Omega m\omega}P_{lm} | 
   D_z
|\Psi_{\Omega m \omega}P_{l'm}\rangle 
= -0.55
\delta_{ll'} 
{~ \rm a.u.}
\label{dopvalue}
\end{equation}
 determining interaction with the external electric field and $J_+^e = J_x^e + iJ_y^e$
\begin{multline}
J_{+}^{e}=
   \langle
   \Psi_{\Omega=1/2m\omega}P_{lm} |J^e_+
|\Psi_{\Omega=-1/2 m \omega-1}P_{l'm}\rangle \\
= 0.992 \delta_{ll'}
\label{pme}
\end{multline}
and
\begin{multline}
e^2Qq_0 =  \langle
   \Psi_{\Omega m\omega}P_{lm} | \\
   e^2 \sum_q (-1)^q \hat{Q}^2_q({\bf \rm I^{\rm Lu}}) \sum_a \sqrt{\frac{2\pi}{5}}\frac {Y_{2q}(\theta_{1a}, \phi_{1a})}{{r_{1a}}^3} \\
   |\Psi_{\Omega m \omega}P_{l'm}\rangle = 
   -5012~ \delta_{ll'} {~ \rm MHz},
\end{multline}
where $Q=3.49$~barn is the quadrupole moment for the $^{175}$Lu nucleus~\cite{Pekka:2008,stone2014}, 
 were calculated in the present work.
To calculate the $\mathcal{T,P}$-odd shifts the average value of corresponding Hamiltonians (\ref{Wd2},\ref{Htp}) were evaluated.

\section{Results}

Electronic structure calculation confirmed the linear equilibrium geometry for LuOH$^+$ with $R=1.930$ \AA, $r=0.954$ \AA, $\theta=0$~ equilibrium values \footnote{We would like to mention again that these parameters are Jacobi coordinates, defined above; $R$ is not the Lu-O distance.}.
The calculated values of the $\mathcal{T,P}$-violation parameters are $E_{\rm eff}=-29.1$~GV/cm, $E_s=-25.7$~kHz. In both cases contribution of triple cluster amplitudes is 2.7\%. This is slightly larger than this contribution to the $W_M$ parameter calculated for LuOH$^+$ in Ref.~\cite{Maison:20a}.

\begin{figure}[h]
\centering
  \includegraphics[width=0.5\textwidth]{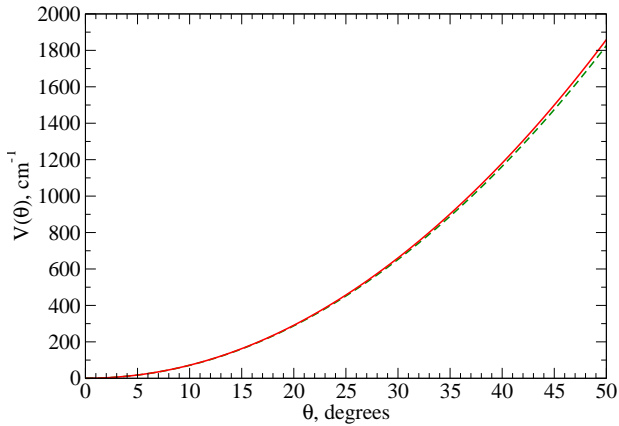}
  \caption{Potential curve $V(\theta)$. Red (solid) curve is for CCSD, green (dashed) curve is for CCSD(T) calculations.}
  \label{curve}
\end{figure}

In Fig.~\ref{curve} and Table~\ref{spec} the calculated potential energy curve and corresponding spectroscopic properties are given. One can see that results for the CCSD and CCSD(T) models are very close to each other. Excitation energy of $\nu_2$ quanta is about 100 cm$^{-1}$ larger than that for the isoelectronic molecule YbOH \cite{Zakharova:22, zhang2021accurate}. The energy difference $\sim 27$cm$^{-1}$ between $\nu_2=2^2$ and $\nu_2=2^0$ states is due to the anharmonicity of the potential and close to that for YbOH \cite{zhang2021accurate}. The $l-$doubling value for $\nu_2=1$ of 23.5 MHz is also close to that for YbOH. Based on our study in Ref.~\cite{Zakharova:22} we estimate the accuracy of the calculation on the level of 10\%. The $l-$doubling for $\nu_2=2^2$ state is about three orders of magnitude smaller. This state can also be used for $\mathcal{CP}$-violation searches and can be completely polarized by electric field of a few V/cm.

In Tables \ref{PE} and  \ref{spec2} the calculated polarizations $P$ and hyperfine energy levels, respectively for the lowest $N=1$ rotational level
of the first excited $v=1$ bending vibrational mode of the $^{175}$LuOH$^+$ for the external electric fields $E=50, 100, 150, 200, 250, 300, 350 $ V/cm are presented. The selected values are comfortable for the experiment and ensure almost saturated values for polarizations.
 The levels are ordered by the energy value.
Here $M_F=M_J+{M^{\rm H}_I}+{M^{\rm Lu}_I}$ is the projection of the total molecular (electronic-rotational-vibrational-nuclear spins) angular momentum ${\bf F}$ on the lab axis. There are 24 levels for $M_F=1/2$ and $M_F=3/2$, 22 levels for $M_F=5/2$, 16 levels for $M_F=7/2$, 8 levels for $M_F=9/2$ and 2 levels for $M_F=11/2$. Calculations showed that all levels have polarizations $P<0.6$. No
level approaches $P=1$ value in accordance to Ref. \cite{Petrov:2022}.

\begin{figure}
\includegraphics[scale=0.29]{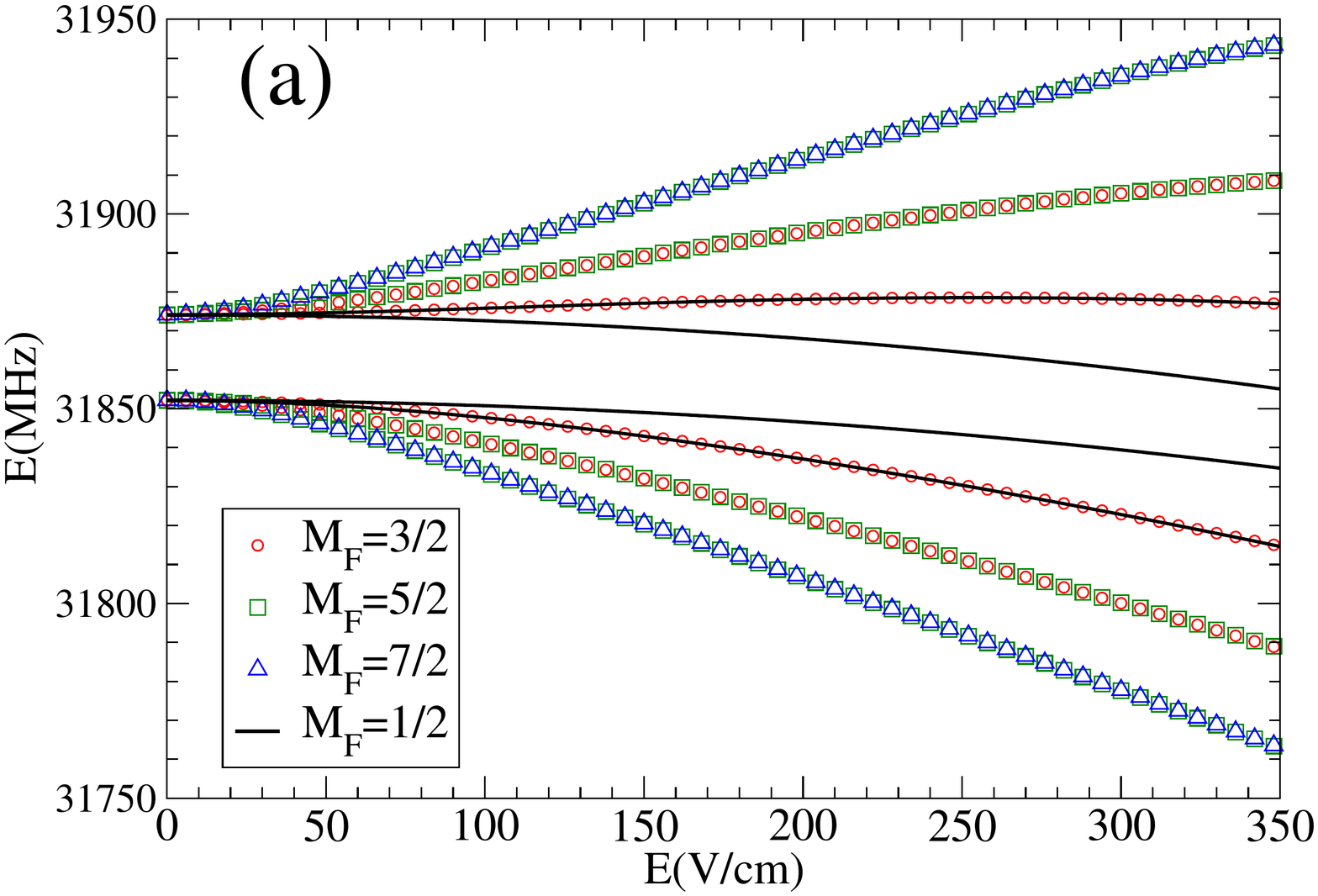}
\includegraphics[scale=0.29]{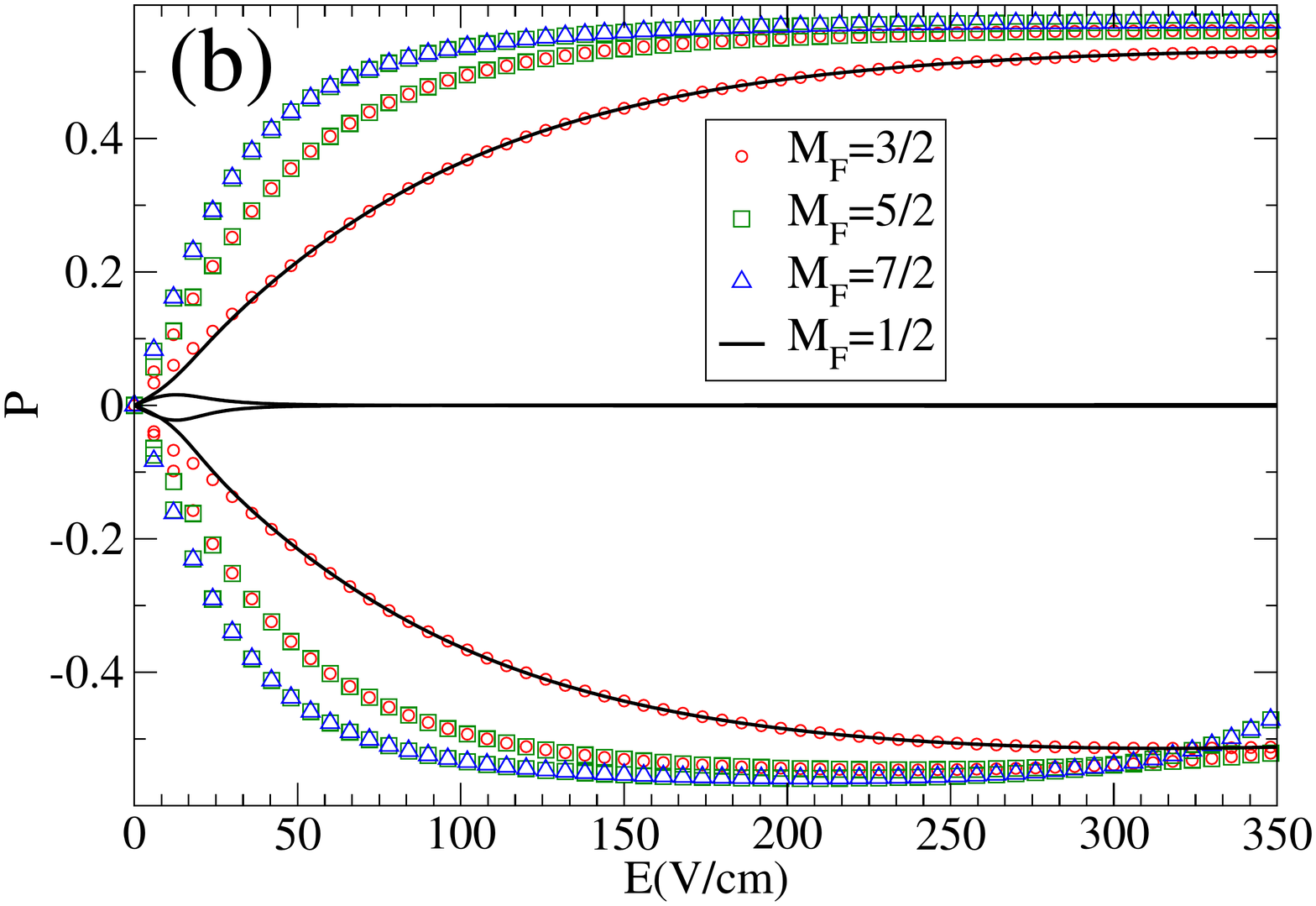}
\caption{\label{EP} (Color online)
  Energy (panel (a)) and polarization (panel(b)) for the group of levels with zero field energy of $\sim$31850 MHz, numbered 43-56.}
\end{figure}

 In Fig. \ref{EP} the calculated energies and $P$ for the group of levels with zero field energy of $\sim$31850 MHz as functions of the external electric field are presented.  The selected states are appropriate for experiment. They have the largest polarizations, and it is the energy grouping with the fewest number of states (only 14). There are almost degenerate states with close values of $P$. These states differ by only projection ${M^{\rm H}_I=\pm 1/2}$ which almost does not influence energy and $P$ due to the weakness of the hyperfine interaction with hydrogen nucleus. Two states with almost zero sensitivity to $e$EDM have close to zero projection of total less hydrogen nuclear spin $M_J+{M^{\rm Lu}_I}$ on the lab axis.

Finally we calculated  vibrational energy levels 
and $l-$doubling effect for the first two bending excitation modes,
hyperfine energy and sensitivity to $e$EDM and the scalar-pseudoscalar nucleus-electron interaction, given by Eq.~(\ref{shift}), for all hyperfine levels of the first excited bending mode of $^{175}$LuOH$^+$. Calculations are required for preparation and interpretation of the experiment on $\mathcal{T, P}$-violation searches on $^{175}$LuOH$^+$.

Since LuOH$^+$ can be created by reacting laser-cooled atomic Lu+ ions with molecules containing an OH group (e.g., water), it is an ideal candidate molecule to use in a quantum logic spectroscopy type experiment, as detailed in \cite{taylor:2022aa}. With this experimental design, it should be possible to obtain a measurement precision of 15 $\mu$Hz in 300 hours of measurement with 12 molecular ions in an external electric field of 32 V/cm. This electric field corresponds to a polarization of 0.35 for co-magnetometer states 44 and 56 with $M_F=5/2,7/2$. These parameters yield a 1$\sigma$ \eEDM~precision of $3\cdot 10^{-30}$ $e \cdot \textrm{cm}$, which matches the precision of the most recent ACME measurement~\cite{ACME:18}.

\begin{table}
\caption{\label{spec} Calculated vibrational energy levels (${\rm cm}^{-1}$) 
and $l-$doubling (MHz) for the $\nu_2=0-2$ quanta of bending excitation modes of $^{175}$LuOH$^+$. Stretching mode $\nu_1$ and ligand mode $\nu_3$ quanta are zero in calculations.}
\begin{ruledtabular}
\begin{tabular}{ccc}
Parameter &   CCSD & CCSD(T)  \\
\hline
$\nu_2=0$ &  0.  &   0. \\
$\nu_2=1$ &  445.  &   442. \\
$\nu_2=2^0$ & 878.  &  871. \\
$\nu_2=2^2$ &  904.  &   898. \\
$l$-doubling $(\nu_2=1)$  & 23.4   & 23.5 \\
$l$-doubling $(\nu_2=2^2)$  & 0.005   & 0.005 \\
\end{tabular}
\end{ruledtabular}
\end{table}

\begin{table*}
\caption{\label{PE}  The calculated polarizations $P$ for the different projections of the total angular momentum $M_F$ of the lowest $N=1$ rotational level
of the first excited the $v=1$ bending vibrational mode of $^{175}$LuOH$^+$ for selected values of the external electric field (in V/cm). Levels are numbered by increasing energy given in Table \ref{spec2}.}
\begin{ruledtabular}
\begin{tabular}{rrrrrrrrr r rrrrrrrrr}
 & & \multicolumn{7}{c}{Electric field}  & & &\multicolumn{7}{c}{Electric field}\\
 \# & $M_F$ &  50.& 100. &   150. &   200.  &  250. &   300. &   350.       && \#  & $M_F$ & 50. &  100. &   150. &   200.  &  250. &   300. &   350.\\
 \cline{1-9} \cline{11-19}
 1   & 1.5 &-0.3811 &-0.4750& -0.5012& -0.5112& -0.5157& -0.5180 &-0.5190   &&   49   &0.5 &  0.0014& -0.0001& -0.0004& -0.0006& -0.0008& -0.0010& -0.0011   \\
 2   & 2.5 &-0.3812 & -0.4750& -0.5012& -0.5112& -0.5157& -0.5180& -0.5190  &&   50   &0.5 & -0.0013&  0.0002&  0.0005&  0.0006&  0.0008&  0.0010&  0.0011   \\
 3   & 0.5 &-0.2436& -0.3809& -0.4436& -0.4735& -0.4889& -0.4971& -0.5014   &&   51   &0.5 & -0.2151& -0.3626& -0.4434& -0.4855& -0.5061& -0.5138& -0.5123   \\
 4   & 1.5 &-0.2450 &-0.3811& -0.4437& -0.4736& -0.4890& -0.4972 &-0.5014   &&   52   & 1.5 &-0.2164 &-0.3625& -0.4432& -0.4853& -0.5060& -0.5137 &-0.5122  \\
 5   & 0.5 &-0.0013& -0.0004& -0.0004& -0.0005& -0.0006& -0.0008& -0.0009   &&   53   & 1.5 &-0.3631 &-0.4904& -0.5307& -0.5443& -0.5461& -0.5389 &-0.5204  \\
 6   & 0.5 & 0.0011&  0.0004&  0.0004&  0.0005&  0.0007&  0.0008&  0.0009   &&   54   & 2.5 & -0.3631 & -0.4903& -0.5306& -0.5442& -0.5460& -0.5388& -0.5203 \\
 7   & 1.5 & 0.2448 & 0.3807&  0.4427&  0.4713&  0.4834&  0.4835 & 0.4697   &&   55   & 2.5 & -0.4457 & -0.5331& -0.5540& -0.5591& -0.5563& -0.5396& -0.4656 \\
 8   & 0.5 & 0.2436&  0.3804&  0.4426&  0.4712&  0.4833&  0.4834&  0.4696   &&   56   & 3.5 & -0.4457 & -0.5330& -0.5539& -0.5590& -0.5562& -0.5395& -0.4656 \\ \cline{10-19} 
 9   & 2.5 & 0.3808 &  0.4741&  0.4994&  0.5059& -0.3943& -0.3964& -0.3968  &&   57   & 4.5 &  0.3530 &  0.4008&  0.4123&  0.4169&  0.4194&  0.4209&  0.4221 \\
10   & 1.5 & 0.3807 & 0.4741&  0.4994&  0.5059&  0.4797& -0.0144 &-0.3131   &&   58   & 5.5 &  0.3528 &  0.4006&  0.4121&  0.4167&  0.4191&  0.4207&  0.4219 \\
 \cline{1-9}
11   & 4.5 & -0.3374 & -0.3830& -0.3938& -0.3980& -0.4001& -0.4013& -0.4022 &&   59   & 3.5 &  0.3266 &  0.3896&  0.4054&  0.4108&  0.4128&  0.4134&  0.4133 \\
12   & 3.5 & -0.3374 & -0.3830& -0.3938& -0.3980& -0.4000& -0.4013& -0.4021 &&   60   & 4.5 &  0.3265 &  0.3894&  0.4052&  0.4106&  0.4126&  0.4132&  0.4131 \\
13   & 3.5 & -0.3040 & -0.3691& -0.3863& -0.3926& -0.3953& -0.3964& -0.3968 &&   61   & 2.5 &  0.2853 &  0.3692&  0.3934&  0.4009&  0.4001&  0.3858&  0.3149 \\
14   & 2.5 &-0.3039 & -0.3691& -0.3862& -0.3925&  0.4790& -0.0129& -0.3131  &&   62   & 3.5 &  0.2853 &  0.3691&  0.3932&  0.4007&  0.4000&  0.3857&  0.3148 \\
15   & 1.5 &-0.2449 &-0.3369& -0.3678& -0.3775& -0.3531&  0.1399 & 0.4377   &&   63   & 1.5 & 0.2203 & 0.3258&  0.3670&  0.3829&  0.3869&  0.3821 & 0.3663  \\
16   & 2.5 &-0.2452 & -0.3370& -0.3679& -0.3775& -0.3533&  0.1384&  0.4377  &&   64   & 2.5 &  0.2205 &  0.3257&  0.3668&  0.3828&  0.3868&  0.3819&  0.3662 \\
17   & 0.5 &-0.1419& -0.2446& -0.3027& -0.3338& -0.3488& -0.3511& -0.3388   &&   65   &0.5 &  0.1202&  0.2201&  0.2839&  0.3216&  0.3419&  0.3503&  0.3499   \\
18   & 1.5 &-0.1446 &-0.2450& -0.3028& -0.3339& -0.3488& -0.3511 &-0.3388   &&   66   & 1.5 & 0.1241 & 0.2203&  0.2837&  0.3213&  0.3417&  0.3501 & 0.3497  \\
19   & 0.5 &-0.0024& -0.0004& -0.0002& -0.0002& -0.0003& -0.0003& -0.0004   &&   67   &0.5 &  0.0036&  0.0002& -0.0003& -0.0005& -0.0006& -0.0008& -0.0009   \\
20   & 0.5 & 0.0020&  0.0003&  0.0002&  0.0002&  0.0003&  0.0003&  0.0004   &&   68   &0.5 & -0.0035& -0.0002&  0.0003&  0.0005&  0.0006&  0.0007&  0.0008   \\
21   & 0.5 & 0.1422&  0.2444&  0.3025&  0.3343&  0.3516&  0.3609&  0.3653   &&   69   &0.5 & -0.1200& -0.2195& -0.2827& -0.3199& -0.3401& -0.3496& -0.3523   \\
22   & 1.5 & 0.1444 & 0.2448&  0.3026&  0.3344&  0.3517&  0.3609 & 0.3653   &&   70   & 1.5 &-0.1238 &-0.2197& -0.2825& -0.3197& -0.3399& -0.3494 &-0.3522  \\
23   & 1.5 & 0.2447 & 0.3364&  0.3673&  0.3794&  0.3843&  0.3858 & 0.3854   &&   71   & 1.5 &-0.2198 &-0.3246& -0.3647& -0.3796& -0.3835& -0.3820 &-0.3773  \\
24   & 2.5 & 0.2450 &  0.3365&  0.3674&  0.3794&  0.3843&  0.3858&  0.3854  &&   72   & 2.5 & -0.2201 & -0.3245& -0.3646& -0.3795& -0.3834& -0.3819& -0.3772 \\
25   & 2.5 & 0.3036 &  0.3683&  0.3850&  0.3907&  0.3927&  0.3929&  0.3921  &&   73   & 2.5 & -0.2847 & -0.3676& -0.3902& -0.3961& -0.3953& -0.3911& -0.3847 \\
26   & 3.5 &  0.3036 &  0.3684&  0.3850&  0.3908&  0.3927&  0.3929&  0.3921 &&   74   & 3.5 & -0.2847 & -0.3674& -0.3901& -0.3959& -0.3952& -0.3909& -0.3845 \\
27   & 4.5 &  0.3369 &  0.3821&  0.3924&  0.3961&  0.3978&  0.3985&  0.3989 &&   75   & 3.5 & -0.3257 & -0.3876& -0.4017& -0.4047& -0.4033& -0.3993& -0.3932 \\
28   & 3.5 &  0.3369 &  0.3821&  0.3924&  0.3961&  0.3977&  0.3985&  0.3989 &&   76   & 4.5 & -0.3256 & -0.3874& -0.4015& -0.4045& -0.4031& -0.3991& -0.3930 \\ 
\cline{1-9}
29   & 2.5 & 0.0456 &  0.0794&  0.1002&  0.1122&  0.1191&  0.1230&  0.1252  &&   77   & 4.5 & -0.3520 & -0.3988& -0.4093& -0.4128& -0.4143& -0.4148& -0.4149 \\
30   & 3.5 &  0.0457 &  0.0794&  0.1001&  0.1122&  0.1191&  0.1230&  0.1252 &&   78   & 5.5 & -0.3518 & -0.3985& -0.4091& -0.4126& -0.4140& -0.4146& -0.4147 \\ \cline{10-19} 
31   & 1.5 & 0.0312 & 0.0585&  0.0794&  0.0946&  0.1052&  0.1127 & 0.1181   &&   79   & 3.5 &  0.0082 &  0.0150&  0.0201&  0.0241&  0.0273&  0.0302&  0.0328 \\
32   & 2.5 & 0.0314 &  0.0585&  0.0794&  0.0946&  0.1052&  0.1127&  0.1181  &&   80   & 4.5 &  0.0083 &  0.0150&  0.0202&  0.0242&  0.0274&  0.0303&  0.0329 \\
33   & 0.5 & 0.0145&  0.0311&  0.0456&  0.0586&  0.0699&  0.0797&  0.0881   &&   81   & 3.5 &  0.0063 &  0.0119&  0.0164&  0.0197&  0.0217&  0.0227&  0.0225 \\
34   & 1.5 & 0.0162 & 0.0315&  0.0458&  0.0587&  0.0700&  0.0798 & 0.0881   &&   82   & 2.5 &  0.0062 &  0.0118&  0.0163&  0.0196&  0.0217&  0.0226&  0.0224 \\
35   & 0.5 & 0.0016&  0.0003&  0.0001&  0.0001&  0.0001&  0.0001&  0.0001   &&   83   &1.5  & 0.0039 & 0.0082&  0.0119&  0.0150&  0.0172&  0.0185 & 0.0189  \\
36   & 0.5 &-0.0063& -0.0032& -0.0011& -0.0005& -0.0003& -0.0002& -0.0002   &&   84   & 2.5 &  0.0044 &  0.0083&  0.0120&  0.0150&  0.0173&  0.0186&  0.0189 \\
37   & 0.5 &-0.0099& -0.0285& -0.0450& -0.0583& -0.0695& -0.0788& -0.0863   &&   85   &0.5 &  0.0014&  0.0040&  0.0063&  0.0085&  0.0106&  0.0125&  0.0142   \\
38   & 1.5 &-0.0195 &-0.0322& -0.0462& -0.0589& -0.0699& -0.0790 &-0.0865   &&   86   & 1.5 & 0.0025 & 0.0043&  0.0065&  0.0086&  0.0107&  0.0126 & 0.0142  \\
39   & 1.5 &-0.0285 &-0.0588& -0.0801& -0.0952& -0.1055& -0.1121 &-0.1163   &&   87   &0.5 &  0.0007&  0.0003&  0.0001&  0.0001&  0.0000&  0.0000&  0.0000   \\
40   & 2.5 &-0.0331 & -0.0593& -0.0803& -0.0953& -0.1055& -0.1129& -0.1273  &&   88   & 3.5 & -0.0056 & -0.0120& -0.0171& -0.0213& -0.0258& -0.0311& -0.0374 \\
41   & 3.5 & -0.0463 & -0.0805& -0.1018& -0.1142& -0.1214& -0.1253& -0.1273 &&   89   & 4.5 & -0.0069 & -0.0126& -0.0172& -0.0214& -0.0259& -0.0312& -0.0375 \\
42   & 2.5 &-0.0450 & -0.0804& -0.1018& -0.1142& -0.1213& -0.1246& -0.1164  &&   90   & 2.5 & -0.0041 & -0.0089& -0.0150& -0.0204& -0.0259& -0.0320& -0.0387 \\
 \cline{1-9}
43   & 2.5 & 0.4474 &  0.5366&  0.5597&  0.5684&  0.5722&  0.5739&  0.5746  &&   91   & 3.5 & -0.0066 & -0.0110& -0.0155& -0.0205& -0.0260& -0.0321& -0.0387 \\
44   & 3.5 &  0.4473 &  0.5365&  0.5597&  0.5683&  0.5721&  0.5739&  0.5746 &&   92   &0.5 & -0.0008& -0.0018& -0.0028& -0.0031& -0.0021& -0.0010& -0.0004  \\
45   & 1.5 & 0.3642 & 0.4928&  0.5350&  0.5515&  0.5586&  0.5613 & 0.5618   &&   93   & 1.5 &-0.0025 &-0.0054& -0.0091& -0.0134& -0.0179& -0.0222 &-0.0260  \\
46   & 2.5 & 0.3642 &  0.4927&  0.5349&  0.5515&  0.5585&  0.5613&  0.5617  &&   94   &0.5 & -0.0010& -0.0021& -0.0037& -0.0067& -0.0116& -0.0171& -0.0223  \\
47   &0.5 &  0.2156&  0.3638&  0.4457&  0.4894&  0.5128&  0.5252&  0.5309   &&   95   & 1.5 &-0.0030 &-0.0061& -0.0094& -0.0132& -0.0180& -0.0238 &-0.0306  \\
48   & 1.5 & 0.2170 & 0.3637&  0.4455&  0.4892&  0.5127&  0.5251 & 0.5308   &&   96   & 2.5 & -0.0049 & -0.0091& -0.0124& -0.0169& -0.0222& -0.0280& -0.0339 

\end{tabular}
\end{ruledtabular}
\end{table*}

\begin{table*}
\caption{\label{spec2}  
The calculated energies (in MHz) for the different projections of the total angular momentum $M_F$  of the lowest $N=1$ rotational  level of the first excited $v=1$ bending vibrational mode of $^{175}$LuOH$^+$ for the selected values of the external electric field (in V/cm). Levels are numbered by the increasing energy. Zero energy level corresponds to the lowest energy of $N=1$ states at zero electric field.
} 
\begin{ruledtabular}
\begin{tabular}{rrrrrrrrr r rrrrrrrrr}
 & & \multicolumn{7}{c}{Electric field}  & & &\multicolumn{7}{c}{Electric field}\\
\#  & $M_F$ & 50. &  100. &   150. &   200.  &  250. &   300. &   350.&& \#   & $M_F$ & 50. &  100. &   150. &   200.  &  250. &   300. &   350.\\
 \cline{1-9} \cline{10-19}
1 &1.5 &   -5 &  -16 &   -28  &  -41  &  -54  &  -68 &   -82    &&    49 & 0.5 &31852 & 31851  &31849  &31847  &31843  &31839  &31835  \\ 
2 &2.5 &   -5 &  -16 &   -28  &  -41  &  -55  &  -68 &   -82    &&    50 & 0.5 &31874 & 31873  &31871  &31868  &31864  &31860  &31855  \\ 
3 & 0.5 &  -2 &    -6  &  -12  &  -19  &  -27  &  -35  &  -44   &&   51 & 0.5 &31875 & 31876  &31877  &31878  &31879  &31878  &31877  \\ 
4 &1.5 &   -2 &   -6 &   -12  &  -19  &  -27  &  -35 &   -44    &&    52 &1.5 &31875 &31876 & 31877  &31878  &31879  &31878 & 31877 \\    
5 & 0.5 &   0 &    -1  &   -2  &   -4  &   -6  &   -9  &  -13   &&   53 &1.5 &31877 &31883 & 31889  &31895  &31901  &31905 & 31909 \\    
6 & 0.5 &  23 &    23  &   21  &   19  &   17  &   14  &   10   &&   54 &2.5 &31877 & 31883 & 31889 & 31895 & 31901 & 31905 & 31909 \\   
7 &1.5 &   25 &   28 &    32  &   35  &   38  &   41 &    43    &&    55 &2.5 &31880 & 31891 & 31903 & 31914 & 31925 & 31935 & 31944 \\   
8 & 0.5 &  25 &    28  &   32  &   35  &   39  &   41  &   43   &&  56 &3.5 &31880 & 31891 & 31903 & 31914 & 31925 & 31936 & 31944 \\ \cline{10-19} 
9 &2.5 &   29 &   39 &    49  &   60  &   69  &   55 &    41    &&    57 &4.5 &32043 & 32027 & 32010 & 31992 & 31974 & 31956 & 31938 \\   
10 &1.5 &   29 &   39 &    49  &   60  &   70  &   75 &    66   &&    58 &5.5 &32043 & 32027 & 32010 & 31992 & 31974 & 31956 & 31939 \\
 \cline{1-9}
11 &4.5 &  119 &   102 &    85 &    67 &    49 &    31 &    13  &&   59 &3.5 &32046 & 32033 & 32019 & 32004 & 31988 & 31973 & 31957 \\   
12 &3.5 &  119 &   102 &    85 &    67 &    49 &    32 &    14  &&   60 &4.5 &32046 & 32033 & 32019 & 32004 & 31988 & 31973 & 31957 \\   
13 &3.5 &  122 &   110 &    97 &    83 &    69 &    55 &    40  &&   61 &2.5 &32048 & 32039 & 32028 & 32016 & 32003 & 31991 & 31980 \\   
14 &2.5 &  122 &  111 &    97  &   84  &   70  &   74 &    66   &&    62 &3.5 &32049 & 32039 & 32028 & 32016 & 32003 & 31991 & 31980 \\   
15 &1.5 &  125 &  118 &   109  &  100  &   90  &   86 &    93   &&    63 &1.5 &32051 &32045 & 32037  &32028  &32019  &32009 & 32000 \\    
16 &2.5 &  125 &  118 &   109  &  100  &   90  &   86 &    93   &&    64 &2.5 &32051 & 32045 & 32037 & 32028 & 32019 & 32009 & 32000 \\   
17 & 0.5 & 128 &   125  &  121  &  116  &  110  &  104  &   99  &&   65 & 0.5 &32052 & 32049  &32045  &32040  &32033  &32027  &32020  \\ 
18 &1.5 &  128 &  125 &   121  &  116  &  110  &  104 &    99   &&    66 &1.5 &32052 &32049 & 32045  &32040  &32034  &32027 & 32020 \\    
19 & 0.5 & 128 &   128  &  127  &  125  &  123  &  121  &  119  &&   67 & 0.5 &32053 & 32051  &32049  &32046  &32043  &32039  &32035  \\ 
20 & 0.5 & 152 &   151  &  150  &  148  &  146  &  144  &  142  &&   68 & 0.5 &32076 & 32074  &32072  &32068  &32064  &32060  &32055  \\ 
21 & 0.5 & 153 &   154  &  156  &  158  &  160  &  162  &  164  &&   69 & 0.5 &32076 & 32076  &32076  &32075  &32074  &32073  &32071  \\ 
22 &1.5 &  153 &  154 &   156  &  158  &  160  &  162 &   164   &&    70 &1.5 &32076 &32076 & 32076  &32075  &32074  &32073 & 32071 \\    
23 &1.5 &  155 &  161 &   168  &  175  &  182  &  189 &   195   &&    71 &1.5 &32078 &32081 & 32085  &32089  &32091  &32094 & 32096 \\    
24 &2.5 &  155 &  161 &   168  &  175  &  182  &  189 &   195   &&    72 &2.5 &32078 & 32081 & 32085 & 32089 & 32091 & 32094 & 32096 \\   
25 &2.5 &  158 &  169 &   182  &  194  &  206  &  218 &   230   &&    73 &2.5 &32080 & 32088 & 32096 & 32103 & 32111 & 32117 & 32123 \\   
26 &3.5 &  158 &   169 &   181 &   194 &   206 &   218 &   230  &&   74 &3.5 &32080 & 32088 & 32096 & 32103 & 32111 & 32117 & 32123 \\   
27 &4.5 &  162 &   178 &   195 &   213 &   231 &   249 &   267  &&   75 &3.5 &32083 & 32095 & 32107 & 32120 & 32132 & 32143 & 32154 \\   
28 &3.5 &  162 &   178 &   196 &   213 &   231 &   249 &   267  &&   76 &4.5 &32083 & 32095 & 32107 & 32120 & 32132 & 32143 & 32154 \\
 \cline{1-9}
29 &2.5 &  689 &  687 &   685  &  682  &  680  &  677 &   674   &&    77 &4.5 &32086 & 32102 & 32119 & 32137 & 32155 & 32172 & 32190 \\   
30 &3.5 &  689 &   687 &   685 &   683 &   680 &   677 &   674  &&   78 &5.5 &32086 & 32102 & 32120 & 32137 & 32155 & 32173 & 32190 \\  \cline{10-19} 
31 &1.5 &  689 &  689 &   690  &  690  &  691  &  692 &   694   &&    79 &3.5 &32320 & 32319 & 32317 & 32316 & 32314 & 32313 & 32311 \\   
32 &2.5 &  689 &  689 &   690  &  690  &  691  &  692 &   694   &&    80 &4.5 &32320 & 32319 & 32317 & 32316 & 32314 & 32312 & 32311 \\   
33 & 0.5 & 690 &   691  &  693  &  695  &  699  &  703  &  707  &&   81 &3.5 &32320 & 32321 & 32322 & 32323 & 32325 & 32328 & 32331 \\   
34 &1.5 &  690 &  691 &   693  &  695  &  699  &  703 &   707   &&    82 &2.5 &32320 & 32321 & 32322 & 32323 & 32325 & 32328 & 32331 \\   
35 & 0.5 & 690 &   691  &  694  &  697  &  701  &  707  &  713  &&   83 &1.5 &32321 &32323 & 32325  &32329  &32334  &32339 & 32346 \\    
36 & 0.5 & 713 &   715  &  717  &  720  &  724  &  729  &  735  &&   84 &2.5 &32321 & 32322 & 32325 & 32329 & 32334 & 32339 & 32346 \\   
37 & 0.5 & 714 &   715  &  717  &  721  &  725  &  730  &  736  &&   85 & 0.5 &32321 & 32323  &32327  &32333  &32339  &32347  &32355  \\ 
38 &1.5 &  713 &  715 &   717  &  721  &  725  &  730 &   736   &&    86 &1.5 &32321 &32323 & 32327  &32333  &32339  &32347 & 32355 \\    
39 &1.5 &  714 &  715 &   718  &  722  &  726  &  732 &   738   &&    87 & 0.5 &32321 & 32324  &32328  &32334  &32341  &32350  &32359  \\ 
40 &2.5 &  714 &  715 &   718  &  722  &  727  &  732 &   737   &&    88 &3.5 &32342 & 32344 & 32347 & 32351 & 32356 & 32362 & 32368 \\   
41 &3.5 &  714 &   716 &   719 &   723 &   727 &   732 &   737  &&   89 &4.5 &32342 & 32344 & 32347 & 32351 & 32356 & 32362 & 32368 \\   
42 &2.5 &  714 &  716 &   719  &  723  &  727  &  732 &   738   &&    90 &2.5 &32342 & 32344 & 32348 & 32353 & 32359 & 32366 & 32374 \\  
 \cline{1-9}
43 &2.5 &31846 &31834 & 31820  &31806  &31792  &31778 & 31763   &&    91 &3.5 &32342 & 32344 & 32348 & 32353 & 32359 & 32366 & 32374 \\   
44 &3.5 &31846 & 31834 & 31821 & 31807 & 31792 & 31778 & 31763  &&   92 & 0.5 &32342 & 32344  &32348  &32353  &32360  &32368  &32376  \\ 
45 &1.5 &31849 &31841 & 31832  &31822  &31811  &31800 & 31788   &&    93 &1.5 &32342 &32344 & 32348  &32353  &32360  &32368 & 32377 \\    
46 &2.5 &31849 &31841 & 31832  &31822  &31811  &31800 & 31789   &&    94 & 0.5 &32342 & 32345  &32348  &32354  &32360  &32368  &32377 \\  
47 & 0.5 &31851 & 31848  &31843  &31837  &31830  &31823  &31815 &&  95 &1.5 &32342 &32345 & 32348  &32354  &32360  &32368 & 32377 \\   
48 &1.5 &31851 &31848 & 31843  &31837  &31830  &31823 & 31815   &&    96 &2.5 &32342 & 32345 & 32348 & 32353 & 32360 & 32368 & 32377             
\end{tabular}
\end{ruledtabular}
\end{table*}

\begin{acknowledgments}
     Electronic structure calculations have been carried out using computing resources of the federal collective usage center Complex for Simulation and Data Processing for Mega-science Facilities at National Research Centre ``Kurchatov Institute'', http://ckp.nrcki.ru/.

    $~~~$Molecular rovibrational structure calculations have been supported by the Russian Science Foundation Grant No. 18-12-00227. Calculations of property integrals were supported by the Foundation for the Advancement of Theoretical Physics and Mathematics ``BASIS'' Grant according to Projects No. 20-1-5-76-1 and No. 21-1-2-47-1.
\end{acknowledgments}

\bibliographystyle{apsrev}

\end{document}